\documentclass[superscriptaddress,twocolumn,prb,preprintnumbers,amsmath,amssymb]{revtex4}
\usepackage{graphicx,afterpage}
\usepackage{xcolor}
\usepackage[colorlinks=true,plainpages=false,linkcolor=blue,urlcolor=blue,citecolor=blue,pdfpagemode=UseNone,pdfstartview=FitBH]
{hyperref}

\begin{document}
\title{NMR study of charge-density waves under hydrostatic pressure in YBa$_2$Cu$_3$O$_{\rm{y}}$}

\author{I. Vinograd}
\email{grvngrd@gmail.com}
\affiliation{Univ. Grenoble Alpes, INSA Toulouse, Univ. Toulouse Paul Sabatier, EMFL, CNRS, LNCMI, 38000 Grenoble, France}
\author{R. Zhou}
\affiliation{Univ. Grenoble Alpes, INSA Toulouse, Univ. Toulouse Paul Sabatier, EMFL, CNRS, LNCMI, 38000 Grenoble, France}
\author{H. Mayaffre}
\affiliation{Univ. Grenoble Alpes, INSA Toulouse, Univ. Toulouse Paul Sabatier, EMFL, CNRS, LNCMI, 38000 Grenoble, France}
\author{S. Kr\"amer}
\affiliation{Univ. Grenoble Alpes, INSA Toulouse, Univ. Toulouse Paul Sabatier, EMFL, CNRS, LNCMI, 38000 Grenoble, France}
\author{R. Liang}
\affiliation{Department of Physics and Astronomy, University of British Columbia, Vancouver, BC, Canada, V6T~1Z1}
\affiliation{Canadian Institute for Advanced Research, Toronto, Canada}
\author{W.N.~Hardy}
\affiliation{Department of Physics and Astronomy, University of British Columbia, Vancouver, BC, Canada, V6T~1Z1}
\affiliation{Canadian Institute for Advanced Research, Toronto, Canada}
\author{D.A.~Bonn}
\affiliation{Department of Physics and Astronomy, University of British Columbia, Vancouver, BC, Canada, V6T~1Z1}
\affiliation{Canadian Institute for Advanced Research, Toronto, Canada}
\author{M.-H. Julien}
\email{marc-henri.julien@lncmi.cnrs.fr}
\affiliation{Univ. Grenoble Alpes, INSA Toulouse, Univ. Toulouse Paul Sabatier, EMFL, CNRS, LNCMI, 38000 Grenoble, France}
\date{\today}


\begin{abstract}


The effect of hydrostatic pressure ($\boldsymbol{P}$) on charge-density waves (CDW) in YBa$_2$Cu$_3$O$_{\rm{y}}$ has recently been controversial. Using nuclear magnetic resonance (NMR), we find that both the short-range CDW in the normal state and the long-range CDW in high fields are, at most, slightly weakened at $\boldsymbol{P} = 1.9$~GPa. This result is in contradiction with x-ray scattering results finding complete suppression of the CDW at $\sim$1~GPa and we discuss possible explanations of this discrepancy. Quantitative analysis, however, shows that the NMR data is not inconsistent with a disappearance of the CDW on a larger pressure scale, typically $\sim$10-20~GPa. We also propose a simple model reconciling transport data with such a hypothesis, provided the pressure-induced change in doping is taken into account. We conclude that it is therefore possible that most of the spectacular increase in $T_c$ upon increasing pressure up to $\sim$15~GPa arises from a concomitant decrease of CDW strength. 
\end{abstract}
\maketitle

\section{Introduction}

\begin{figure*}
\hspace*{-1cm}   
  \includegraphics[width=13.5cm]{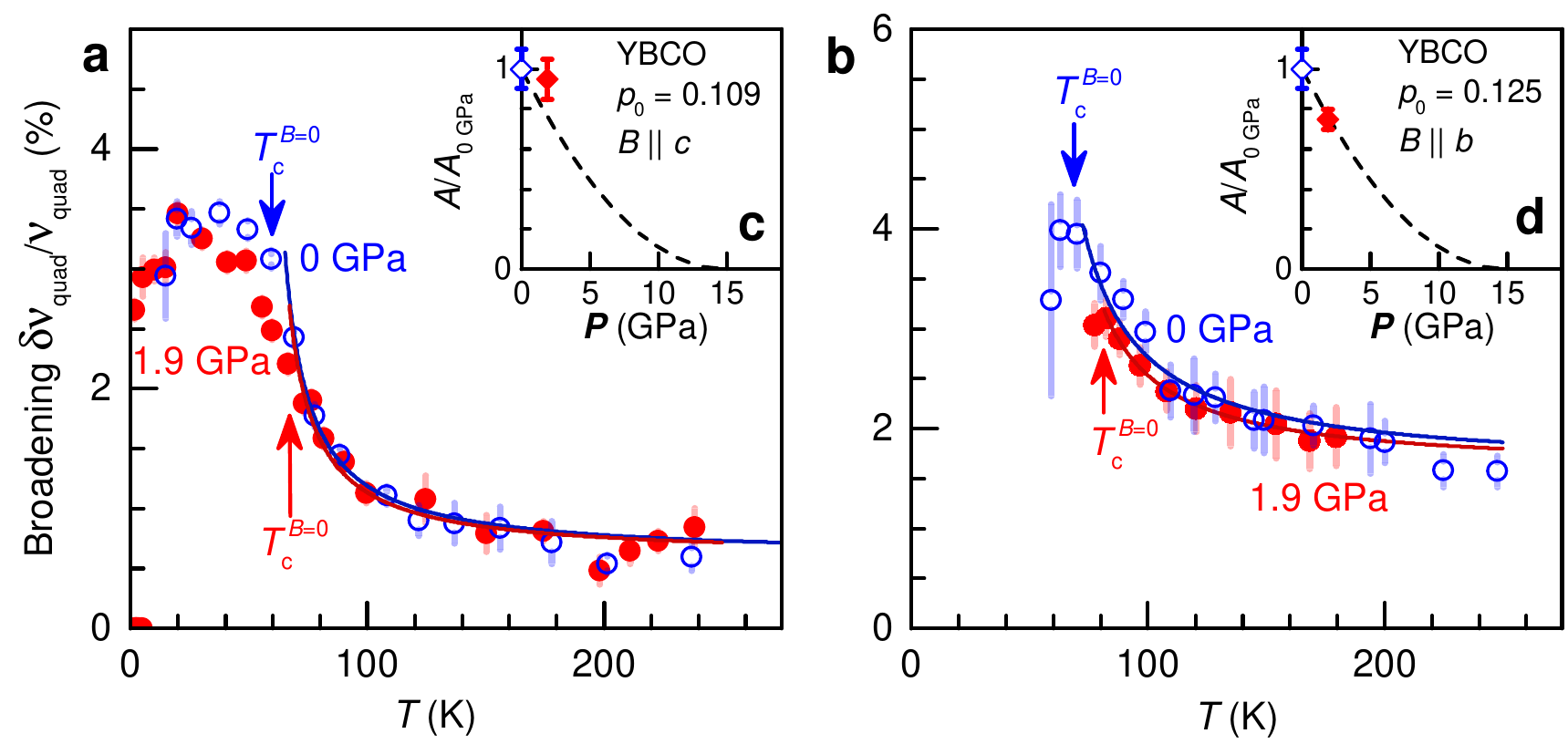}
  \caption{\label{broadening} (a) Quadrupole broadening $\delta\nu_{\rm{quad}}/\nu_{\rm{quad}}$ (due to short-range CDW order), at 0~GPa and 1.9~GPa for the O-II sample (O(2) site and field tilted by 16-18$^\circ$ from the $c$-axis). Below $T_{\rm c}$ the two data sets differ since superconductivity is favored under pressure, at the expense of charge order. If data had been taken in the same field, the difference  below $T_c$ would be even larger ($B=10$~T at 0~GPa, $B=15$~T at 1.9~GPa, with a few points at 12~T above $T_c$). (b) Quadrupole broadening in O-VIII (O(3F) site, $B\parallel b=15$~T). Lines are Curie-Weiss fits (see text) sharing the same background $C=0.59$\% in (a) and $C=1.58$\% in (b). Insets (c) and (d) show the pressure dependence of the amplitude $A$ in Curie-Weiss fits. The dashed lines correspond not to a fit but to a quadratic dependence on $\boldsymbol{P}$ vanishing at $\boldsymbol{P}_{c}=15$~GPa (see text). In (c), that $A(\boldsymbol{P})$ misses the data point at 1.9~GPa for O-II may be ascribed to an increase in doping that partially compensates the intrinsic suppression of the CDW at this pressure.}
\end{figure*}

High-$T_c$ superconductivity in the cuprates arises in close proximity to a charge-density wave (CDW) phase. A challenge in the field is to understand how both phenomena compete and whether, behind pure competition, there is a more involved relationship between them. To tackle this question, experiments in YBa$_2$Cu$_3$O$_{\rm{y}}$ have used temperature, magnetic field, hole-doping or uniaxial strain as tuning parameters~\cite{Wu2011,Wu2013,Gerber2015,Chang2016,Jang2016,Ghiringhelli2012,Achkar2012,Chang2012,Huecker2014,Blanco-Canosa2014,LeTacon2014,Forgan2015,Wu2015,Julien2015,Kim2018}. The effect of hydrostatic pressure, on the other hand, is controversial. 

The application of a 15~GPa hydrostatic pressure in underdoped YBa$_2$Cu$_3$O$_{6.6}$ results in an increase of $T_{\rm c}$ from $\sim$64~K to 107~K~\citep{Sadewasser2000}, which is significantly higher than $T_{\rm c}= 94$~K of optimally-doped YBa$_2$Cu$_3$O$_{\rm{y}}$ at ambient pressure. This has long remained a mystery but Cyr-Choini\`{e}re \textit{et al.} have recently remarked that the sensitivity of $T_{\rm c}$ to pressure correlates with the strength of CDW order in YBa$_2$Cu$_3$O$_{\rm{y}}$~\citep{Cyr-Choiniere2018}. They have thus suggested that, because charge order competes with superconductivity, it is the suppression of the CDW phase under pressure that actually drives the $T_{\rm c}$ increase.

This proposal has been challenged by two sets of experiments in YBa$_2$Cu$_3$O$_{\rm{y}}$ (YBCO) that, however, appear to be mutually contradictory. On the one hand, two x-ray studies have found that pressures as small as 1~GPa are sufficient to fully suppress signatures of short-range charge order~\citep{Souliou2018,Huang2018}. Such a rapid suppression is thus inconsistent with a link between CDW and the increase of $T_{\rm c}$ up to $\sim$15~GPa. On the other hand, pressures of $\sim$1~GPa hardly affect two prominent signatures of charge order in transport measurements, namely slow quantum oscillations and a negative Hall number $R_{\rm{H}}$~\citep{Putzke2016,Putzke2018,Cyr-Choiniere2018}. From this observation, Putzke {\it et al.}~\cite{Putzke2018}  have concluded that the $T_c$ increase under pressure and the $T_c$ depression near $p=0.12$ at ambient pressure~\cite{Liang2006} are both unrelated to the CDW. Thus, three incompatible viewpoints have been expressed: the pressure dependence of charge order is either too weak~\citep{Putzke2016,Putzke2018}, too strong~\citep{Souliou2018,Huang2018} or of the right magnitude~\citep{Cyr-Choiniere2018} to explain the rise in $T_c$ up to 15~GPa. In contrast, an NMR study argues that pressure actually enhances charge order in YBa$_2$Cu$_3$O$_{6.9}$~\cite{Reichardt2018}.

Since transport, but not scattering, experiments have been performed in high magnetic fields to suppress superconductivity and since high fields are known to strengthen charge order, the apparent conflict between transport and scattering measurements might be explained if pressure affects the short-range CDW observed in zero-field~\cite{Ghiringhelli2012,Achkar2012,Chang2012,Wu2015} but not the long-range CDW in high fields~\cite{Wu2011,Wu2013,Gerber2015,Julien2015,Chang2016,Jang2016}. This explanation would however question the widespread belief that high-field transport properties reflect a Fermi-surface reconstruction by the short-range 2D CDW~\cite{Harrison2011,Allais2014,Zhou2017PNAS,Laliberte2018}. Therefore, resolving these contradictions is important for elucidating the effect of pressure but also, more broadly, for understanding the CDW. 
 
In this article, we report $^{17}$O nuclear magnetic resonance (NMR) experiments under pressure in high quality, YBa$_2$Cu$_3$O$_{\rm{y}}$ untwinned single crystals. Using a clamp-type cell (see Appendix A and B for experimental details), we applied a pressure of 1.9~GPa to two crystals with ortho-II (O-II) chain-oxygen order (hole doping level $p_0=0.109$, $T_c=59.8$~K) and ortho-VIII (O-VIII) order ($p_0=0.125$, $T_c=67.8$~K), both used in our previous works~\cite{Wu2013,Wu2015,Wu2016,Zhou2017PNAS,Zhou2017PRL,Kacmarcik2018}. One of our main results is to show that the discrepancy between transport and x-rays is unrelated to the field dependence of CDW phases in YBa$_2$Cu$_3$O$_{\rm{y}}$ as the amplitude of both the short-range and the long-range CDW orders are found in NMR to be, at most, weakly affected by a pressure of 1.9~GPa. 

The paper is organized as follows: we first present data concerning the short-range CDW in the normal state (which is field independent) and discuss the discrepancy with x-ray scattering results. We then present data concerning the long-range CDW in high fields. In the last part of the paper, we discuss quantitative aspects of the normal state results, we evaluate the pressure-induced increase in doping and we propose an alternative interpretation of the transport results of Putzke {\it et al.}~\cite{Putzke2018}. Appendices also contain details about the estimation of the pressure-induced increase in hole-doping.

\section{Short-range CDW: experimental results}

Short-range CDW order in YBa$_2$Cu$_3$O$_{\rm{y}}$ produces a spatial modulation of the electric field gradient (EFG) at planar Cu and O sites that leads to a quadrupolar line broadening~\cite{Wu2015}. We find that the dimensionless quadrupole broadening $\frac{\delta\nu_{\rm{quad}}}{\nu_{\rm{quad}}}$, where $\nu_{\rm{quad}}$ is the separation between adjacent quadrupole satellites, is essentially unaffected by a pressure of 1.9~GPa for both samples, at least at temperatures  ($T$) for which superconductivity is absent (Figs. \ref{broadening}a,b). At this stage, we already reach our first important conclusion: 1.9~GPa is not sufficient to completely suppress the CDW at neither $p_0=0.109$ nor 0.125 doping.


\begin{figure*}
\hspace*{-1cm}   
  \includegraphics[width=14cm]{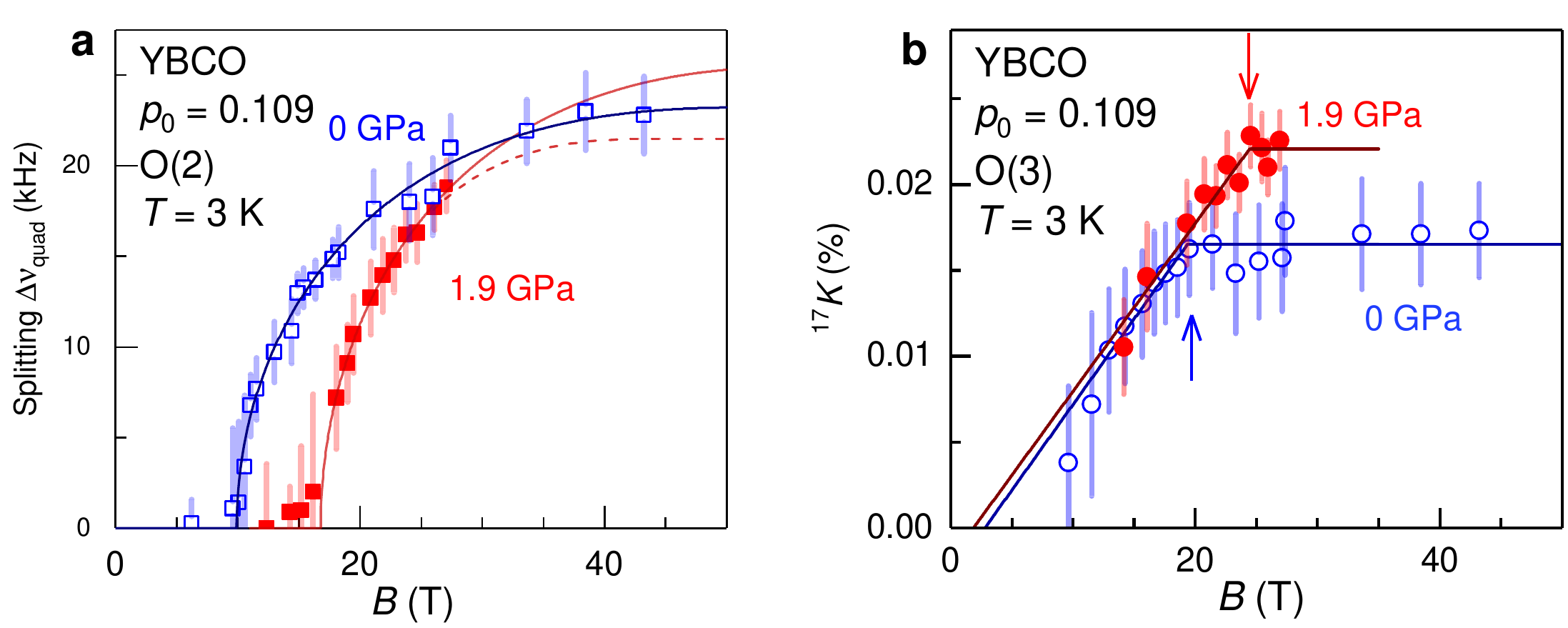}
  \caption{\label{shift}(a) Field dependence of the quadrupole splitting of O(2) sites (due to long-range CDW order), at $T=3$~K for 0~GPa and 1.9 GPa. 0~GPa data are from a re-analysis of the data in ref.~\citep{Wu2013} (see Appendix A). $B$ values correspond to the $c$-axis component of the applied field. Lines are fits to Eq.~\ref{eq:tanh fit}, with the dashed red line corresponding to a fit excluding the highest field point to visualise the uncertainty in the saturation value $\Delta \nu_{\rm{quad}}^{max}$.(b) Field dependence of $^{17}K$ of O(3) sites in the O-II sample at $T=3$~K. Arrows at 19.4$\pm$2~T (0~GPa) and 24.4$\pm$2~T (1.9~GPa) mark the saturation field $B_{\rm sat}$ of $^{17}K$, suggesting a 5$\pm$3~T increase of the upper critical field $B_{\rm c2}$ under pressure ($B_{\rm sat}\simeq B_{\rm c2}$ at low $T$~\cite{Zhou2017PNAS}). Lines are fits according to the procedure described in refs.~\cite{Zhou2017PNAS,Kacmarcik2018}.}
\end{figure*}

\section{Comparison with x-ray scattering}

We now comment on the discrepancy between our NMR and the transport measurements on one side and the x-ray studies in YBCO on the other side. First, a hard x-ray diffraction measurement at $p=0.13$ finds no CDW intensity already at 1~GPa~\citep{Huang2018}. However, as no ambient pressure measurements were performed inside the diamond anvil pressure cell (DAC), it is unclear whether the sensitivity is sufficient to detect the weak CDW signal within the cell~\citep{Huang2018}. Second, an inelastic x-ray scattering study at $p=0.12$~\citep{Souliou2018} found that two signatures of CDW order in an acoustic phonon branch, namely a broadening on cooling followed by a sudden narrowing and partial softening below $T_{\rm{c}}$~\cite{LeTacon2014}, both disappear between 0.8 and 1.5~GPa.

As to the phonon broadening, we remark that the measurements under pressure were performed at a single temperature, near $T_c$. However, the narrowing below $T_c$ is so abrupt~\cite{LeTacon2014} that even a minor misevaluation of $T_c$ under pressure can result in measurements being inadvertently performed slightly below $T_c$ or in the transition region where fluctuations effects and/or sample inhomogeneities may play a role. The phonon softening, on the other hand, is probably a more solid piece of evidence so questioning this result would challenge the interpretation of scattering experiments. One possibility is that the complicated CDW structure factor~\cite{Forgan2015} changes with pressure. Another hypothesis we would like to raise is that the softening of the phonon in question is, one way or another, associated with in-phase ($l=1$) CDW correlations. Even though experiments under strain rather suggest an anticorrelation between this acoustical phonon and long-range 3D CDW~\cite{Kim2018}, in both cases there is a phonon softening along the $b$-axis. The relatively intricate situation and the nearness of transition temperatures at $p \simeq 0.12$ doping ($T_{\rm{CDW}}^{\rm 3D}\simeq 50$~K~\citep{Wu2011,Wu2013,Gerber2015,Laliberte2018} {\it vs.} $T_c\simeq$~65~K) call for further investigation of this question.
 
In principle, other factors may lead to divergence between NMR/transport and x-rays. First, the hydrostaticity of the oils used as pressure medium in the transport and NMR measurements is not as good as that of helium used for both x-ray studies. However, pressure has been applied at room $T$ where the oil is still liquid so non-hydrostaticity is only expected from strains if the solidification with cooling is inhomogeneous. It seems unlikely that small shear strains will dominate over high, largely isotropic pressures since YBCO has a relatively large bulk modulus of about 120~GPa~ \cite{Lei1993}. Furthermore, uniaxial pressure has a very anisotropic effect on $T_c$~\cite{Kraut1993} so if non-hydrostaticity was significant, the pressure-induced change in $T_c$ would be different for different pressure media, which is not the case.


Also, the CDW that we see under pressure cannot be ascribed to pressure-induced disorder in our sample because we do not see any line broadening at $\rm{CuO}$-chain sites. On the contrary, our preliminary data (Appendix D) are consistent with slightly lower disorder under pressure, as reported by Huang {\it et al.}~\cite{Huang2018}. Finally, it is possible that NMR probes preferentially the fully static, pinned, CDW modulations while scattering experiments also integrate fluctuations. However, this goes against the scaling between NMR and x-ray data at ambient pressure~\cite{Wu2015}.

\section{Long-range CDW: experimental results}

So far, we have discussed the impact of pressure on short-range CDW because of its direct relevance to the conflict between transport and x-ray results. We have however pursued our NMR investigation of the CDW into high fields, which has not been probed by scattering under pressure yet.

The NMR signature of the long-range CDW phase is a quadrupole splitting $\Delta \nu_{\rm quad}$ of the lines~\cite{Wu2011,Wu2013,Zhou2017PRL}. $^{17}$O lineshapes in the O-II sample were found to be similar at 0 and 1.9 GPa, suggesting that the CDW wave vector is unchanged. $\Delta \nu_{\rm quad}$ values were obtained at both pressures by fitting O(2) satellites with a set of two peaks (see Appendix A). The onset field of long-range CDW order $B_{\rm CDW}$ is determined by fitting $\Delta \nu_{\rm{quad}}(B)$ to
\begin{equation}\label{eq:tanh fit}
 \Delta \nu_{\rm{quad}}^{max}\tanh\Bigg(1.74\sqrt{\frac{B_{\rm{sat}}-B_{\rm{CDW}}}{B_{\rm{sat}}-B}-1}\Bigg),
\end{equation}
where the field dependence of $\Delta \nu_{\rm{quad}}$ is analogous to the $T$ dependence of a superconducting BCS gap. The very good fit to the 0~GPa data obtained with Eq.~\ref{eq:tanh fit} suggests that the CDW amplitude still increases somewhat above the superconducting upper critical field $B_{c2}(T=0)\simeq 24$~T~\cite{Zhou2017PNAS,Grissonnanche14} before eventually saturating at $B_{\rm{sat}}=54$~T, in contrast with the suggested saturation of the Knight shift. This could indicate that CDW order still competes strongly with superconducting fluctuations. However, we point out that we use Eq.~\ref{eq:tanh fit} without any theoretical justification, mostly for determining the onset field of long-range CDW order, $B_{\rm CDW}$. Within error bars, it is possible that $\Delta \nu_{\rm{quad}}$  saturates above $B_{c2}$.

As Fig.~\ref{shift}a shows, the main effect of increasing pressure is to shift $B_{\rm CDW}$ from 9.9~T to $\sim$16.8~T, {\it i.e.} by $\sim$3.6~T/GPa. Concomitantly, $B_{\rm{c2}}$ also increases, as suggested by our $^{17}K(B)$ data in Fig.~\ref{shift}b, which are consistent with $\text{d}B_{\rm{c2}}/\text{d}{\boldsymbol P} \approx$~3~T/GPa deduced from the irreversibility field~\citep{Putzke2018}. That $\text{d}B_{\rm{c2}}/\text{d}{\boldsymbol P} \approx \text{d}B_{\rm{CDW}}/\text{d}{\boldsymbol P}$ highlights the intimate connection between superconductivity and the field-induced CDW transition CDW suggested in previous works~\cite{Wu2013,Jang2016,Laliberte2018}. Here, the increase of $B_{\rm CDW}$ suggests that pressure has reduced the spatial extension of the CDW halos nucleated in vortex cores~\cite{Wu2013}.

Because of the increased $B_{\rm CDW}$ and $B_{\rm{c2}}$, fields below 30~T are not sufficient to reach a saturation of $\Delta \nu_{\rm quad}$ and thus there remains uncertainty as to whether the amplitude of the charge modulation at $B \gtrsim B_{\rm c2}$ changes with pressure (Fig.~\ref{shift}a). Extrapolation of the fits to higher fields suggests that the amplitude does not change by more than $\pm10\%$, which is in line with the absence of a strong change in the strength of the short-range CDW at this $p_0=0.109$ doping.

\section{Quantitative aspects of the normal-state results and the importance of doping change}

We now come back to the results of Fig.~\ref{broadening} and discuss more quantitative aspects. This part of the paper is more speculative for two main reasons: 1) pressure-induced changes in various quantities are relatively small at 1.9~GPa compared to experimental uncertainties, 2) our proposed interpretation relies on a number of assumptions (essentially linear approximations for pressure-induced changes). However, we shall argue that these assumptions are reasonable and, furthermore, regardless of the degree of uncertainty in the propositions below, our attempt at a quantitative description has the merit of highlighting effects that have been overlooked in some of the previous works. We have deliberately separated this part from the presentation of the results in section II in order to emphasize that the central conclusion of this work, namely that short-range CDW order is still present at 1.9~GPa, is disconnected from the quantitative interpretation. 

Close inspection of the data suggests that $\frac{\delta\nu_{\rm{quad}}}{\nu_{\rm{quad}}}$ is slightly but systematically reduced at 1.9~GPa in the O-VIII sample (Fig. \ref{broadening}b), unlike in the O-II sample (Fig. \ref{broadening}a), even though the experimental error bars are somewhat larger than the difference between datasets with and without pressure. Strikingly, the very same dichotomy (namely, no visible change for O-II, small but noticeable change for O-VIII) is also present in the Hall effect results of ref.~\cite{Cyr-Choiniere2018}. This strongly suggests that the CDW is more resilient to pressure in O-II. 

A natural reason for the contrast between these two concentrations is that $p_0=0.125$ (O-VIII) is at the maximum of the dome of the CDW, while $p_0\simeq 0.11$ lies below where the CDW strength is weaker~\cite{Huecker2014,Blanco-Canosa2014}. Therefore, a small pressure-induced increase in doping (expected from the reduced distance between chains and planes) will strengthen the CDW for the O-II sample, while weakly reducing it for O-VIII. If, concomitantly, there is an intrinsic (not doping-related) decrease of the CDW strength due to pressure, the intrinsic and doping effects will compensate at low pressure for O-II whereas they will both act to weaken the CDW for O-VIII, whatever the pressure strength~\cite{Cyr-Choiniere2018}.

A quantitative analysis of the Hall effect data from ref.~\cite{Putzke2018} provides further support for such a compensation effect: for $p_0\simeq 0.11$ (O-II), Putzke \textit{et al.} have found that $T_{0}$, the temperature at which the Hall number $R_{\rm{H}}$ changes its sign because CDW order reconstructs the Fermi surface, varies slowly as a function of pressure: $\frac{\text{d}T_{0}}{\text{d}\boldsymbol{P}}=-1.1$~K/GPa up to 2.6~GPa \citep{Putzke2018}. They concluded that this rate of suppression of the CDW is too weak to explain the concomitant increase of $T_{\rm{c}}(\boldsymbol{P})$ of $\frac{\text{d}T_{\rm{c}}}{\text{d}\boldsymbol{P}}$=+3.8~K/GPa. However, we point out that pressure-induced doping, although very small, can affect the pressure-dependence of $T_{0}$. Indeed, the doping dependence of $T_{0}$ is very strong at ambient pressure: $\sim$1640~K/hole from $p=0.08$ to $p=0.12$~\cite{LeBoeuf2011}. Multiplying this number by our estimation (see Appendix C) of a pressure-induced doping of $\sim$0.0015~holes/GPa leads to an extrinsic increase $\frac{\text{d}T_{0}}{\text{d}\boldsymbol{P}}_{\rm{dop.}}$=+2.5~K/GPa. The experimentally-determined slope $\frac{\text{d}T_{0}}{\text{d}\boldsymbol{P}}_{\rm{tot.}}$ being the combination of a positive doping and a negative intrinsic effect:
\begin{equation} 
\label{eq:extrinsic intrinsic}
\frac{\text{d}T_{0}}{\text{d}\boldsymbol{P}}_{\rm{tot.}} =\frac{\text{d}T_{0}}{\text{d}\boldsymbol{P}}_{\rm{dop.}} +\, \frac{\text{d}T_{0}}{\text{d}\boldsymbol{P}}_{\rm{intr.}} ,
\end{equation}
it follows that $\frac{\text{d}T_{0}}{\text{d}\boldsymbol{P}}_{\rm{intr.}}=-3.6$~K/GPa. Remarkably, this intrinsic suppression of $T_{0}$ (thus of the CDW) is of nearly equal magnitude (but opposite sign) as the rate of +3.8~K/GPa at which $T_{\rm{c}}$ increases. For $p_0=0.125$, $\frac{\text{d}T_{0}}{\text{d}\boldsymbol{P}}_{\rm{dop.}}\simeq 0$ because $\frac{\text{d}T_{0}}{\text{d}p}\simeq0$ at the maximum~\cite{LeBoeuf2011}, so we expect $\frac{\text{d}T_{0}}{\text{d}\boldsymbol{P}}_{\rm{tot.}}=-3.6$~K/GPa, close to the observed value of $-3.3 \pm 1$~K/GPa at $p_0=0.120$~\cite{Cyr-Choiniere2018}.

Furthermore, if one assumes that $\frac{\text{d}T_{0}}{\text{d}\boldsymbol{P}}_{\rm{intr.}}$ is independent of $\boldsymbol{P}$ one can estimate $T_{0}(\boldsymbol{P})$ to first order from the relation: 
\begin{equation}
\label{eq:T0 vs P}
T_{0}(\boldsymbol{P}) = T_{0}(p_0+\Delta p)+\, \frac{\text{d}T_{0}}{\text{d}\boldsymbol{P}}_{\rm{intr.}}\cdot \boldsymbol{P}  
\end{equation}
The first term accounts for the doping effect and is given by the nearly-parabolic doping dependence of $T_{0}$ at 0~GPa. Following ref.~\citep{Cyr-Choiniere2018}, the doping change under pressure $\Delta p$ is taken to be proportional to both the applied pressure $\boldsymbol{P}$ and the initial doping $p_0$: $\Delta p =  b\cdot p_{0}\cdot \boldsymbol{P}$. In Appendix C, we justify the choice of the value $b=0.014$. 

 \begin{figure}
\hspace*{-1cm}   
  \includegraphics[width=6.75cm]{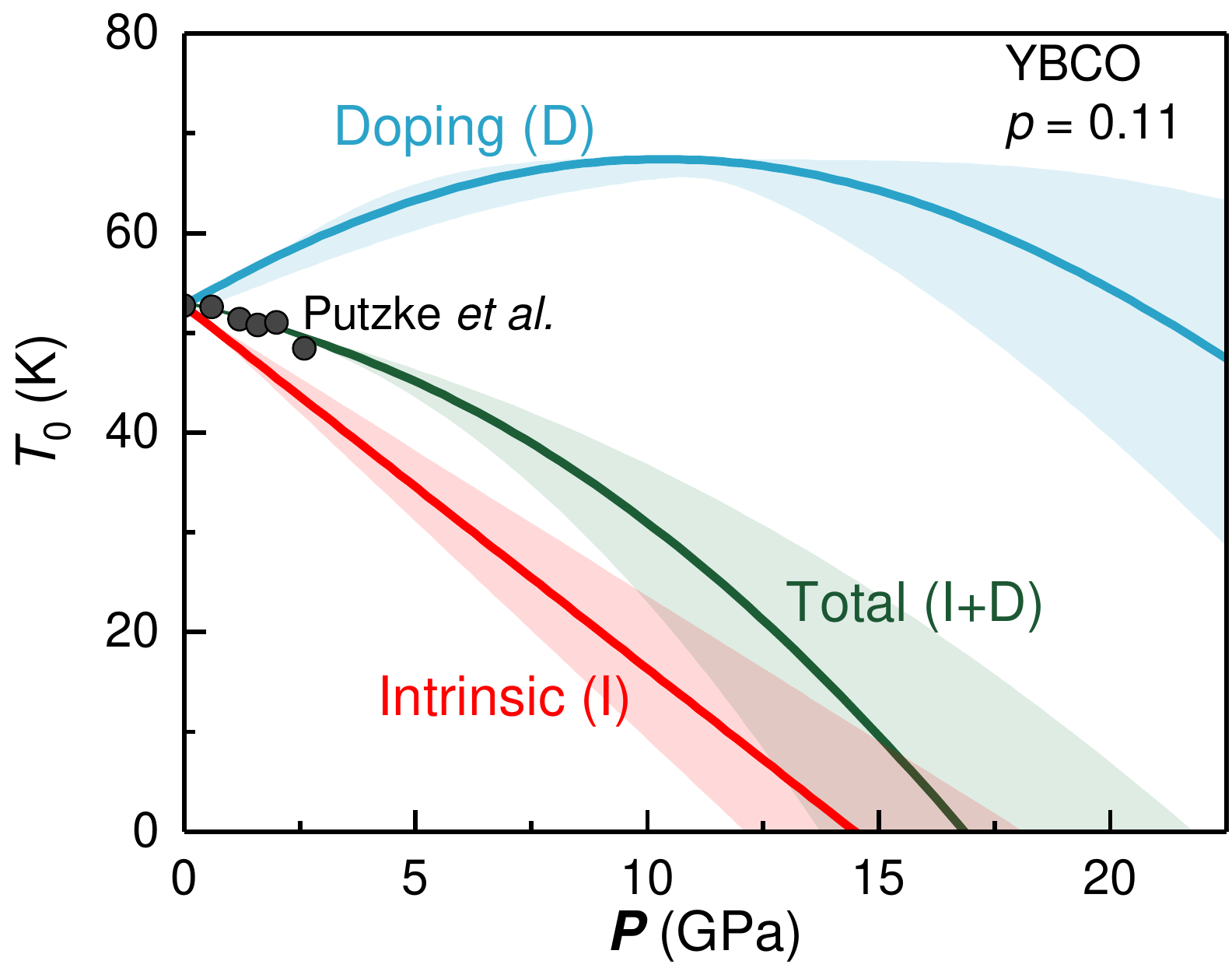}
    \caption[$T_{0}$ \& $T_{\rm{c}}$ pressure dependence]{\label{T0}Calculated pressure dependence of $T_{0}$ for the $p_0= 0.111$ YBCO sample studied by Putzke \textit{et al.} \citep{Putzke2018}. The green curve includes both doping effects (blue) and intrinsic pressure effects (red) on $T_{0}$ and is calculated from Eq.~\ref{eq:T0 vs P} and a 4$^{\rm{th}}$-order polynomial fit of the $T_{0}(p)_{\rm{0\,GPa}}$ data of ref.~\cite{LeBoeuf2011}. The initial slope is equal to the measured $\frac{\text{d}T_{0}}{\text{d}\boldsymbol{P}}_{\rm{tot.}}$=-1.1 K/GPa, as found in ref.~\citep{Putzke2018} and is assumed to be $\boldsymbol{P}$-independent for simplicity. Thick lines are calculated with $b=0.014$ in Eq.~\ref{eq:T0 vs P} and the shaded areas are delimited by results of the calculation with $b=0.010$ and $b=0.018$ (see text).}
\end{figure}

For $p_0 \simeq 0.11$, Eq.~\ref{eq:T0 vs P} predicts complete suppression of $T_{0}$, and hence of charge order, at $\boldsymbol{P}_{c} \sim 17$~GPa~(Fig.~\ref{T0}), comparable to the pressure at which $T_{\rm{c}}$ appears to saturate~\citep{Sadewasser2000}. Thus, once the pressure-induced doping is considered at $p_0\simeq0.11$, we see that: 1) the increase in doping may partially compensate intrinsic effects of the pressure on the CDW below $\sim$10~GPa, 2) opposed to the conclusions by Putzke \textit{et al.}~\citep{Putzke2018}, it is actually possible that the suppression of the CDW goes hand in hand with the pressure-induced increase of $T_{\rm{c}}$, as originally proposed by Cyr-Choini\`{e}re \textit{et al.} \citep{Cyr-Choiniere2018}.

Putting aside the special case of O-II, we now focus our quantitative analysis of NMR data on the O-VIII sample. Since the growth of short-range CDW order does not follow the $T$ dependence of a typical order-parameter, we fit the data above $T_c$ with a Curie-Weiss type dependence: $\frac{\delta\nu_{\rm{quad}}}{\nu_{\rm{quad}}} = \frac{A}{T-\theta} + C$. $A$ is related to the CDW-amplitude, $\theta$ can be seen as the temperature at which the CDW susceptibility would diverge if superconductivity did not intervene and $C$ represents the $T$ independent broadening due to chemical and lattice inhomogeneities as well as unresolved inequivalent sites when the field is tilted off the $c$-axis (subtle crystallographic differences related to the oxygen-ordered structure). If we assume that $C$ is independent of ${\boldsymbol P}$, the fitting indicates that $A$ decreases by 25$\pm12$\% between 0 and 1.9~GPa (Fig.~\ref{broadening}b), with a concomitant change of $\theta$ from $49 \pm 5$~K to $54 \pm 2$~K. With only two data points, there is obviously significant freedom to describe the $\boldsymbol{P}$ dependence of $A$. However, we point out that a quadratic dependence $A(\boldsymbol{P}) \propto (\boldsymbol{P}_{c}-\boldsymbol{P})^2$ vanishing at $\boldsymbol{P}_{c}=$~15~GPa is consistent with the data (inset to Fig.~\ref{broadening}b). Such a dependence is expected if the atomic displacements $u$ are linear in $\boldsymbol{P}$. Indeed, $\frac{\delta\nu_{\rm{quad}}}{\nu_{\rm{quad}}}$ scales with the x-ray scattering intensity~\cite{Wu2015} that, in canonical CDW systems, is proportional to $u^2$. Evidently, the data are not inconsistent with an absence of change within error bars (especially as $C$ may slightly change with ${\boldsymbol P}$ if oxygen order is affected, see Appendix C) but our point here is again that a gradual vanishing of the CDW on a scale of $\sim$15~GPa is also consistent with either the transport or the NMR data. At 1.9~GPa, our estimated doping change $\Delta p \simeq 0.003$ holes~(Appendix C) is too  small to result in a visible change of the CDW amplitude \citep{Huecker2014}. Thus, the decrease of the quadrupole broadening, {\it i.e.} of the CDW amplitude, for the $p_0=0.125$ (O-VIII) sample must be an intrinsic effect of hydrostatic pressure.

\section{Conclusion}

In summary, our NMR data in YBCO show that a pressure of 1.9~GPa has a relatively modest effect, if any, on the strength of each of the two CDW phases. Unlike x-ray scattering studies we do not find a complete suppression of the short-range CDW above 1~GPa. We have discussed the discrepancy between NMR and scattering results and we suggest several experiments to shed light on this issue: verifying whether the CDW peak can be seen by hard x-rays in DACs at ambient pressure, measuring the phonon broadening at higher $T$ under pressure, measuring the phonon softening in a magnetic field (perhaps with neutron scattering), at lower doping and/or with a finer temperature resolution or repeating the NMR experiment with $^{4}$He as pressure medium to perfectly replicate the conditions of the x-ray measurements. We have clarified, and found to be very reasonable, the conditions under which NMR and transport data may be consistent with the original proposal by Cyr-Choini\`{e}re \textit{et al.} that most (about 70\% according to our estimation in Appendix C) of the $T_c$ increase under pressure is due to concomitant weakening of the CDW. This, together with work on stripe order in La$_{2-x}$Ba$_x$CuO$_4$~\cite{Hucker2010}, suggests that hydrostatic pressure may be a convenient, generic tuning parameter of the competition between CDW order and superconductivity in the cuprates.

\section*{Acknowledgements}

We thank S. Badoux, A. Carrington, D. LeBoeuf, T. Klein, M. Le Tacon, C. Proust, A. Sacuto, M. Souliou and S. Wu for helpful discussions as well as M. Hirata invaluable help and advice. 

Work in Grenoble was supported by the Laboratoire d'excellence LANEF in Grenoble (ANR-10-LABX-51-01). Part of this work was performed at the LNCMI, a member of the European Magnetic Field Laboratory (EMFL). Work in Vancouver was supported by the Canadian Institute for Advanced Research and the Natural Science and Engineering Research Council.

\newpage

\appendix

\section{NMR methods}

We used home-built NMR spectrometers and probes, superconducting magnets for fields up to 20~T and the LNCMI M10 resistive magnet for higher fields.

For the O-II sample, the magnetic field $B$ was tilted off the $c$ axis by an angle of 18$^{\circ}$ towards the $b$-axis, as a compromise between resolution of the different O sites (that is maximal for $H\parallel ab$) and a large field component along the $c$-axis, which is required for inducing long-range CDW order~\cite{Wu2011,Wu2013}. For O-VIII, $B$ was applied parallel to the $b$-axis, allowing optimal site separation but hindering investigation of the high-field phase in this configuration. Field values in Figs.~\ref{shift}a,b correspond to the $c$-axis projection of $B$. Values quoted elsewhere are total $B$ values. 

The reference for Knight shift measurements was $^{27}K$ of a metallic Al-foil~\cite{Meissner2013} for the O-II sample($p_0=0.109$) and $^{63}K$ of the Cu-coil~\citep{Kitagawa2010} for the O-VIII sample($p_0=0.125$). 

In the high-field CDW phase, the quadrupole splitting $\Delta \nu_{\rm quad}$ was obtained by fitting the quadrupole satellites by a set of two asymmetric peaks of area ratio of 2:1 as in ref.~\cite{Zhou2017PRL}. The ambient-pressure data in Fig.~\ref{shift}(a) are slightly different from those in ref.~\cite{Wu2013} where the area ratio was not fixed to 2:1 but let free in the fit. This difference has a negligible impact on the determination of the onset field $B_{CDW}$.

\begin{figure}[t!]
\hspace*{-1cm}   
  \includegraphics[width=7cm]{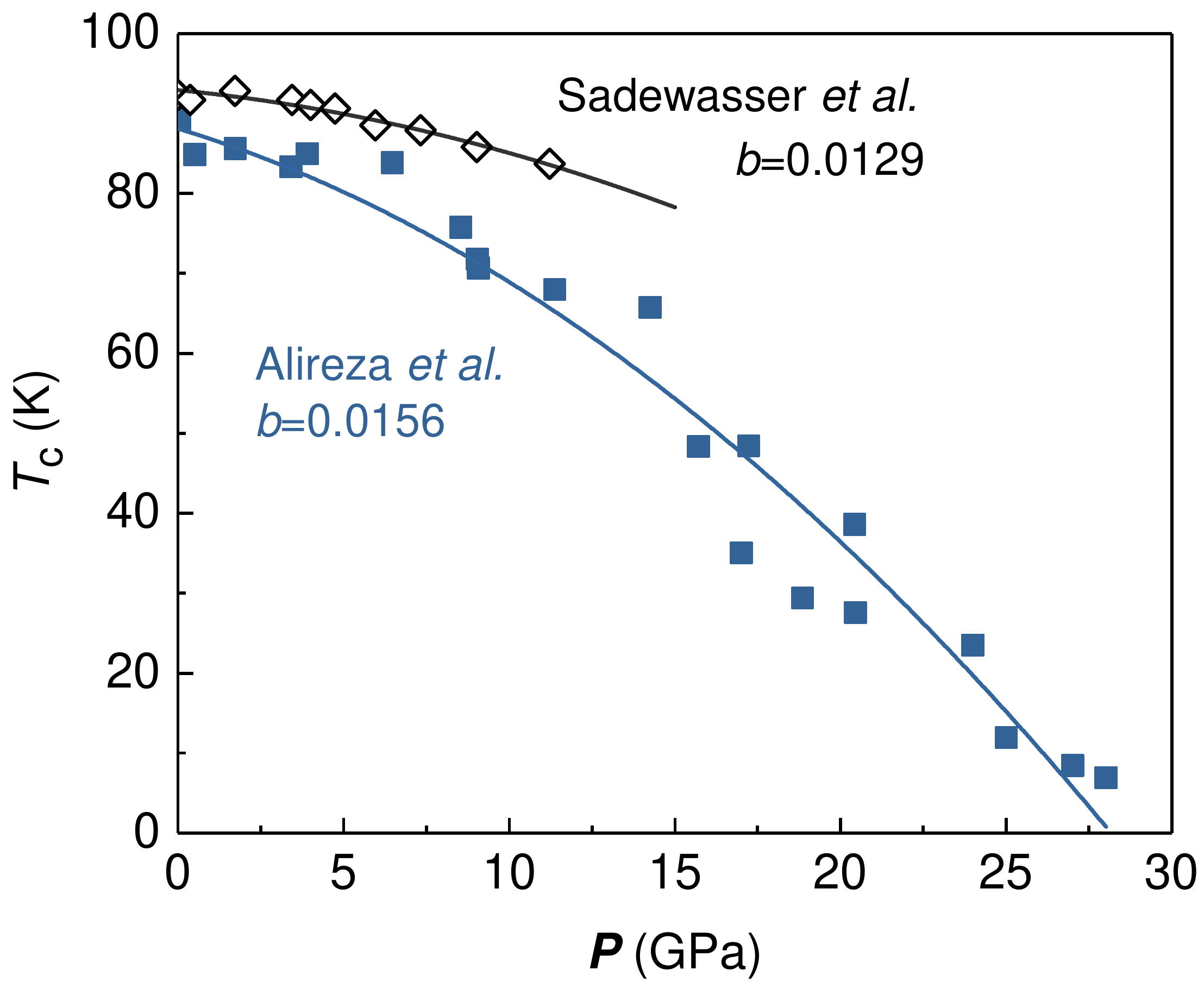}
  \caption{$T_{\rm{c}}(\boldsymbol{P})$ for overdoped YBCO samples of Sadewasser~\textit{et al.} (open diamonds) with $p_0 \approx 0.172$~\citep{Sadewasser2000} and Alireza~\textit{et al.} (filled squares) with $p_0 \approx 0.188$~\citep{Alireza2017}. We fit both the single crystal and polycrystalline samples simultaneously, as their pressure dependence is similar.}
\label{Alireza}
\end{figure}     

\section{Pressure methods}

We used a commercial BeCu/NiCrAl clamp cell from C\&T Factory Co. Ltd. (Japan) and Daphne oil 7373  as a transmitting medium~\citep{Yokogawa2007}. The applied pressure has been calibrated by the resistivity of a long Manganin wire at ambient temperature.

By monitoring the resonance frequency of the NMR tank circuit upon cooling in zero field, we found that $T_{\rm c}$ at 1.9~GPa had increased by 6~K and 13~K for the O-II and O-VIII samples, respectively. These values are in good agreement with data from ref.~\citep{Cyr-Choiniere2018}. This means that, as expected from the specifications of Daphne 7373 at 1.9~GPa~\citep{Yokogawa2007}, no pressure has been lost between 285~K and the low temperature ($T$) range where the oil has solidified. The observed increase of the long-range CDW onset field under pressure (see text) is another, indirect, confirmation of the pressure at low $T$.

The samples were cooled below 250~K within less than two hours after pressurization in order to minimize oxygen reordering in the chains~\cite{Sadewasser2000}. No difference in the NMR properties of the samples could be detected before and after pressurization, so there appears to be no irreversible change after the application of 1.9~GPa.

\section{Pressure-induced doping}

\begin{figure}[t!]
\hspace*{-1cm}   
  \includegraphics[width=8cm]{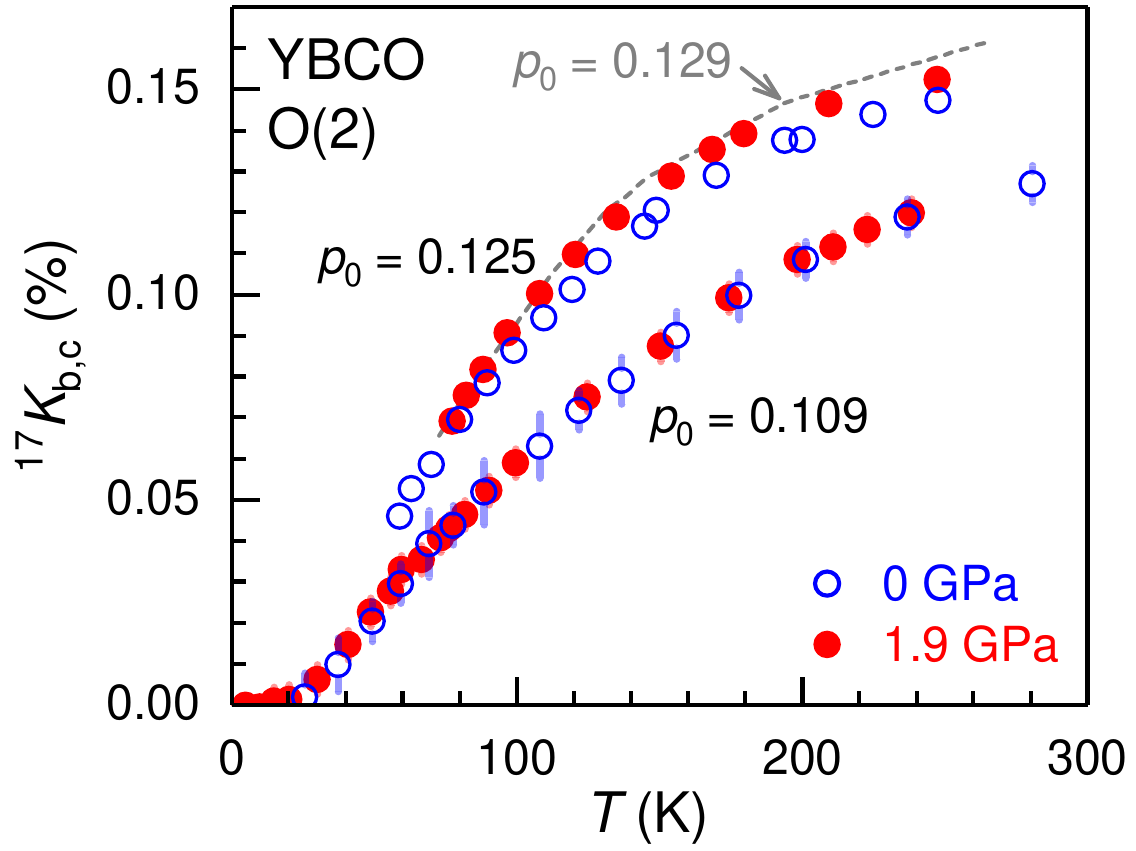}
  \caption{\label{KshiftvsT}$T$-dependence of the O(2) Knight shift $^{17}K$ for the O-II sample ($p_0=0.109$, $B$ tilted by 16-18$^\circ$ from the $c$-axis) and for the O-VIII sample ($p_0=0.125$, $B\parallel b$), both at 0~GPa (blue open symbols) and 1.9~GPa (red closed symbols). Grey dashes: $^{17}K$ for an O-III sample ($p=0.129$) at 0~GPa and $B\parallel b$, shown for comparison.}
\end{figure}

\subsection{Model}

Pressure reduces the distance between ${\mathrm{CuO}}$-chains and ${\mathrm{CuO}}_{\mathrm{2}}$-planes, which facilitates charge transfer and thus increases the hole content $p$. Since the strength of charge order is strongly $p$ dependent, pressure must have a 'doping effect' on the CDW. This needs to be taken into account before discussing quantitatively any possible 'intrinsic effect' of pressure on the CDW. Cyr-Choini\`{e}re \textit{et al.} have proposed that the pressure-induced doping $\Delta p$ is proportional to both the applied pressure $\boldsymbol{P}$ and the initial doping $p_0$~\citep{Cyr-Choiniere2018}:
\begin{equation}\label{eq:induced doping}
\Delta p  =  b\cdot p_{0}\cdot \boldsymbol{P},
\end{equation} 
where $b$ represents the percentage by which the doping increases per GPa. 

Provided that Eq.~\ref{eq:induced doping} is valid, the pressure-induced doping is fully determined by the parameter $b$. Below, we show how the $b$ value can be determined from $T_c(\boldsymbol{P})$ of overdoped samples and from the Knight shift of our $p_0=0.125$ sample.  

\begin{figure}[t!] 
  \includegraphics[width=8.3cm]{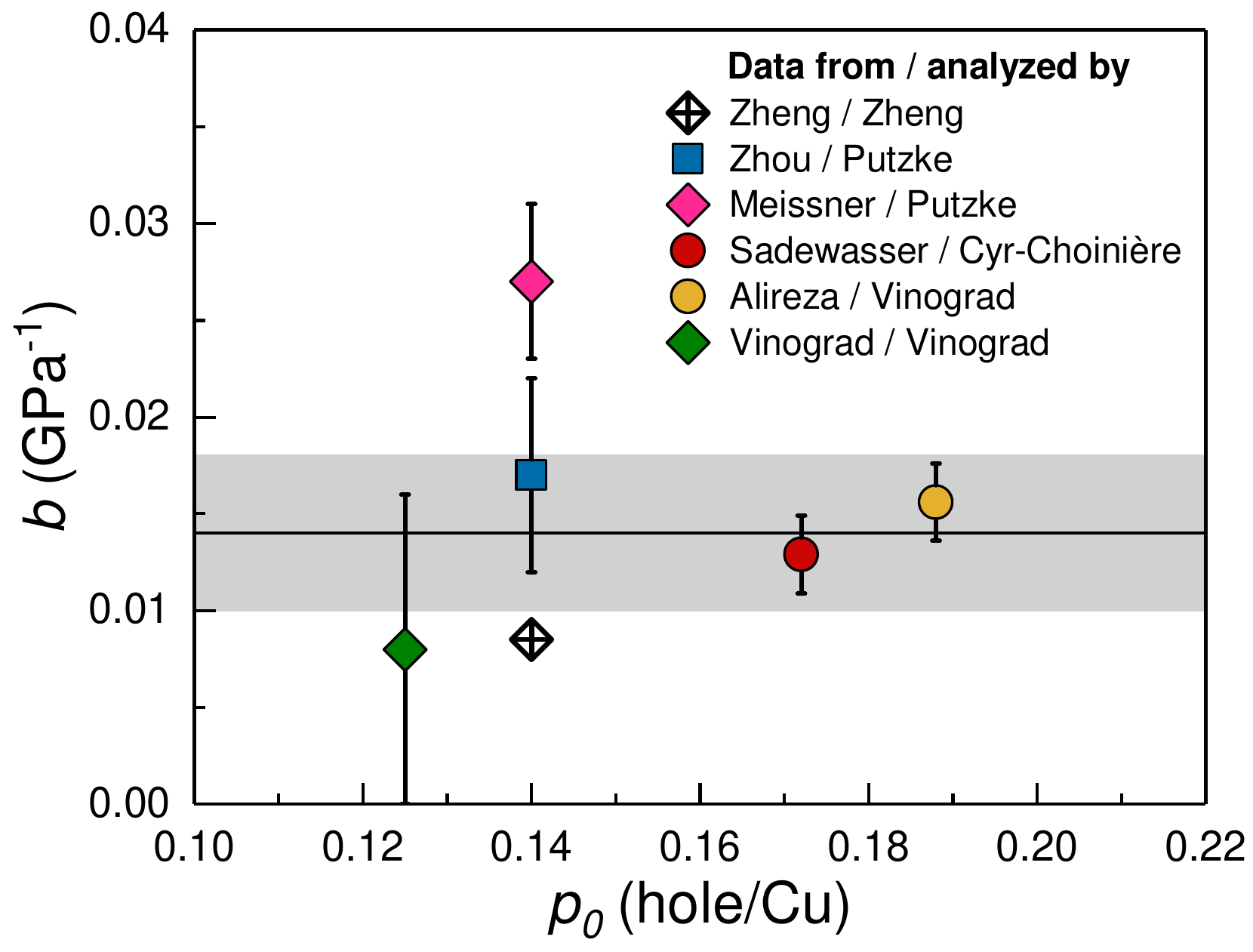}
  \caption{$b$ values (as defined by Eq.~\ref{eq:induced doping}) from Vinograd {\it et al.} (present work), Zheng {\it et al.}~\cite{Zheng1995}, Putzke {\it et al.}~\cite{Putzke2016}, Cyr-Choini\`ere {\it et al.}~\cite{Cyr-Choiniere2018}, Zhou {\it et al.}~\cite{Zhou1996}, Meissner {\it et al.}~\cite{Haase2011}, Sadewasser {\it et al.}~\cite{Sadewasser2000}, Alireza {\it et al.}~\cite{Alireza2017}. The grey shaded bar marks the error of $\pm0.004$~GPa$^{-1}$ in $b$, also used for Fig.~\ref{Tc}. The pink diamond based on $^{17}K$ data of Meissner~\textit{et al.} lies outside of this error. It has been determined from scaling of the full $T$ dependent $^{17}K$ data to a single gap, the pseudogap, whose doping dependence is known. However, $^{17}K$ at low temperature could be partially gapped by the CDW. As the CDW can be intrinsically weakened by increasing $\boldsymbol{P}$ ascribing the full $\boldsymbol{P}$ dependence to the pseudogap alone possibly led to an overestimated increase of the doping.}
  \label{b}
\end{figure}

\subsection{Estimating the pressure-induced doping in the overdoped regime}

Based on Eq.~\ref{eq:induced doping}, a parabolic form of $T_{\rm{c}}(p)$ \citep{Tallon1995} leads to

\begin{equation}\label{eq:Tc vs P}
T_c(p(\boldsymbol{P}))= 94.3 (1-82.6(p_{0}(1+b \cdot \boldsymbol{P})-0.16)^2).
\end{equation} 

By fitting $T_c (\boldsymbol{P})$ data from ref.~\citep{Sadewasser2000} to Eq.~\ref{eq:Tc vs P} (data from overdoped samples must be used as there should be no electronic order affecting $T_c$ in this region of the phase diagram), Cyr-Choini\`{e}re \textit{et al.} determined $b=0.01=1\%$~GPa$^{-1}$ (the exact value used below is 0.0129). As shown in Fig.~\ref{Alireza}, repeating the same procedure with the data of ref.~\cite{Alireza2017}, we find $b=0.0156\,$GPa$^{-1}$. So, from data in overdoped samples, the average value is $b=0.014\,$GPa$^{-1}$. Below, we argue that this value is also representative for underdoped samples. 

\begin{figure}[t!] 
  \includegraphics[width=7cm]{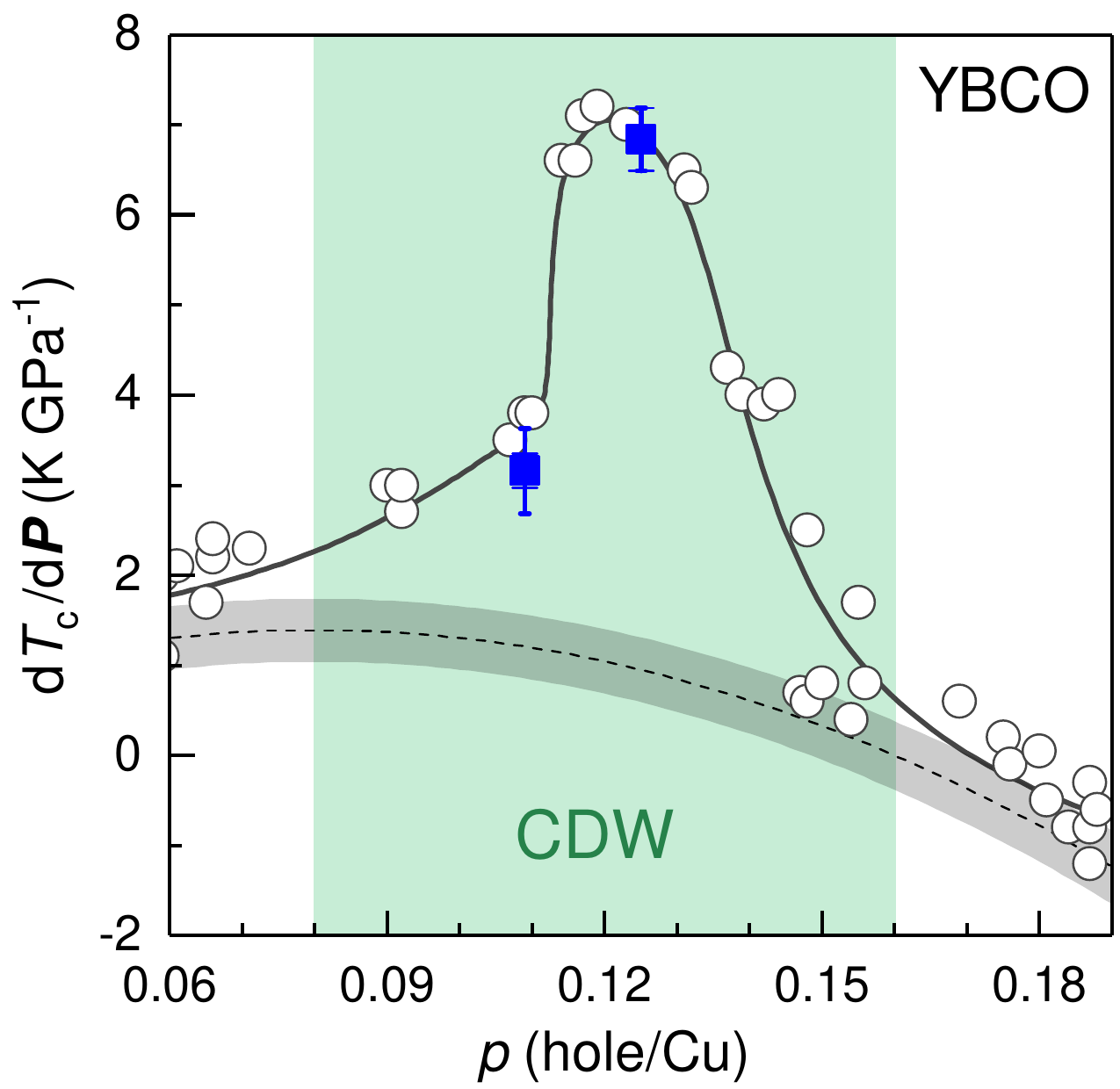}
  \caption{$\text{d}T_c/\text{d}\boldsymbol{P}$ data from Cyr-Choini\'ere {\it et al.}~\citep{Cyr-Choiniere2018} (open circles) compared with the values deduced from the $T_c$ changes of 6~K and 13~K for our O-II and O-VIII samples (blue squares), respectively, at 1.9~GPa (as measured in-situ by the shift in resonance frequency of the NMR tank circuit upon cooling in zero field). The dotted line shows $\text{d}T_c/\text{d}\boldsymbol{P}$ which is calculated from the derivative of the parabolic $T_{\rm{c}}(p)$ using $b=0.014$. The shaded area is bounded by results of the calculation with $b=0.010$ and $b=0.018$.}
  \label{Tc}
\end{figure}

\subsection{Estimating the pressure-induced doping from NMR in underdoped YBa$_2$Cu$_3$O$_y$}

We found a small change in the quadrupole frequency $\nu_Q$  under pressure (from 363 to 368~kHz for O(2) sites in O-II and from 947 to 953~kHz for O(3F) sites in O-VIII) but this result is difficult to interpret because it arises from changes in both the charge density and the lattice parameters. 

The Knight shift $K$, on the other hand, is known to increase monotonously upon increasing $p$ for any $T>200$~K where there should be no contribution from the CDW. As Fig.~\ref{KshiftvsT} shows, a pressure of 1.9~GPa slightly increases $^{17}K$ of the $p_0=0.125$ sample (O-VIII), by about half of the difference with $^{17}K$ in an O-III ($p_0=0.129$) sample at 0~GPa. Thus, assuming that all of the change in $K$ is due to a doping change, $p$ has increased by $(0.129-0.125)/2\simeq 0.002\pm 0.001$ holes at 1.9~GPa. This translates into $b=0.008 \pm 0.008\,$GPa$^{-1}$, which is within error bars consistent with $b=0.014\,$GPa$^{-1}$. 

For the O-II sample, on the other hand, there is no discernible change in $^{17}K$ even though with $b=0.014$, 1.9~GPa should increase $p$ from 0.109 to 0.112. It is possible that the model is too simplified. For instance, details of oxygen ordering could play a role in the charge transfer. At any rate, our data show that the change in doping is very small at 1.9~GPa and thus contributes only weakly to the increase in $T_c$.

\subsection{Pressure-induced doping: summary}

Fig.~\ref{b} summarizes the above determined $b$ values together with other values from the literature. For $b=0.014$ and an initial doping $p_0=0.117$~\cite{Sadewasser2000}, $\Delta p \simeq 0.025$ at 15~GPa, which implies that $T_c$ should raise only to $\sim$76~K, not 107~K, as can be determined from Eq.~\ref{eq:Tc vs P}, if the doping change was the sole effect. This means that only $\sim$30\% of the $T_c$ increase is due to a change in doping. Of course, there remains a large uncertainty on these numbers, given the scattering of the data points shown in Fig.~\ref{b}.

\subsection{Estimating the sensitivity of $T_c$ to pressure}

Using the derivative of the parabolic $T_{\rm{c}}(p)$ we can calculate $\text{d}T_c/\text{d}\boldsymbol{P}$ for a given $b$ value and plot it together with the experimental data. As shown in Fig.~\ref{Tc}, clearly, most of the change of $T_c$ around $p\sim 0.12$ is not due to pressure-induced doping. The calculated change of $T_c$ is much smaller than the experimentally determined $\text{d}T_c/\text{d}\boldsymbol{P}$. However, the calculated values match quite well the experimental data near $p\simeq 0.06$ and $p\simeq 0.18$ where CDW correlations are expected to be negligible.

\section{Effect of pressure on chain order}

\begin{figure}[t!]
  \includegraphics[width=8cm]{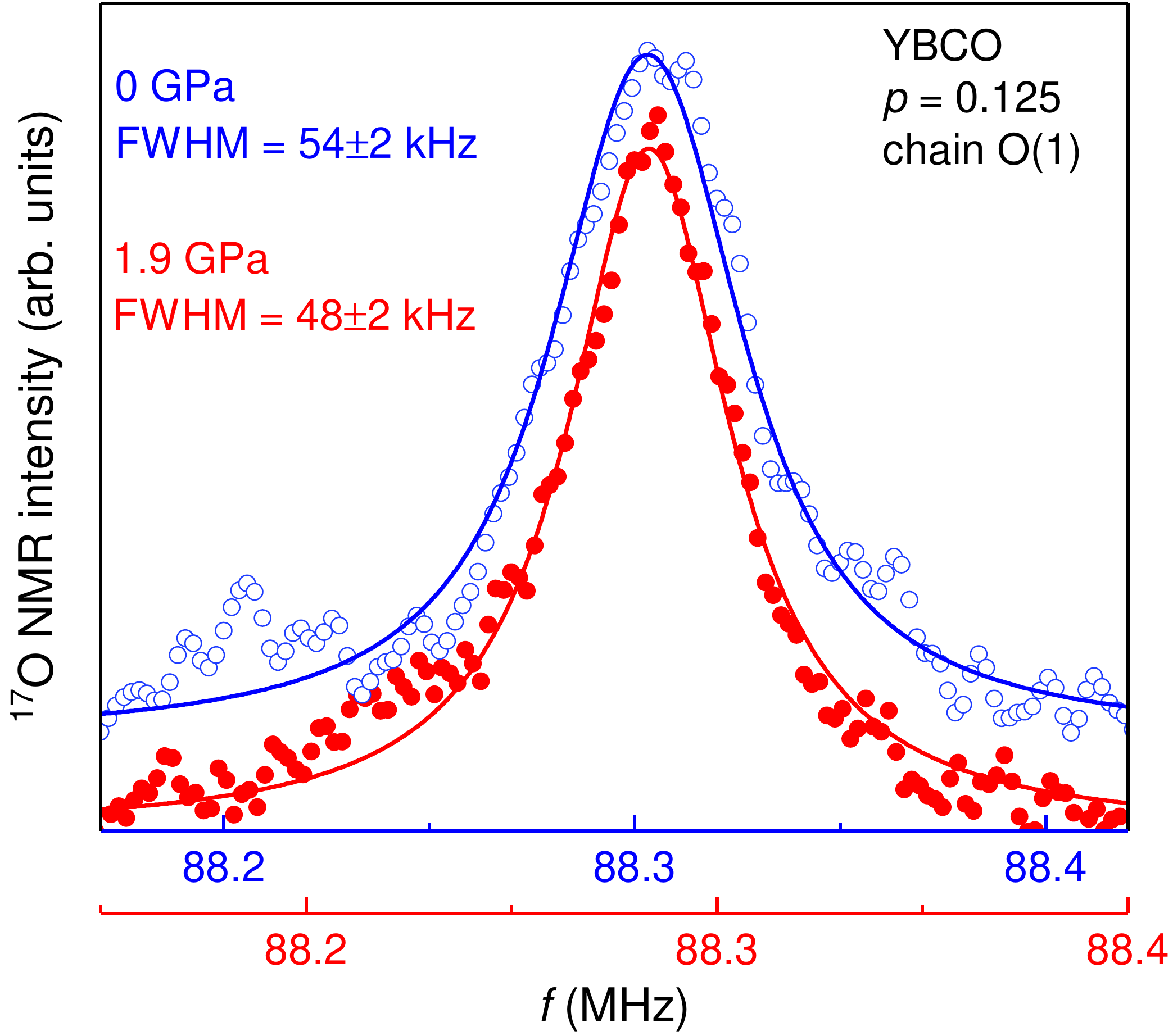}
  \caption{\label{chain}Comparison of first quadrupole satellites of chain O(1) site at 0 and 1.9~GPa, measured at $T=168.5$~K (spectra are offset vertically for clarity). The Lorentzian full-width at half maximum (FWHM) is indicated for each line.} 
\end{figure}

We found a modest, but reproducible, narrowing of oxygen-empty Cu(1E) sites in another O-II sample and oxygen-filled O(1) sites (Fig.~\ref{chain}), which is consistent with a slight increase of oxygen order under pressure, also found in a recent x-ray experiment~\cite{Huang2018}. However, since the values are close to our experimental uncertainty, more precise investigation of this interesting issue would require crystals with larger $^{17}$O concentration on the chain site ({\it i.e.} final annealing under $^{17}$O atmosphere).

\end{document}